\documentclass[fleqn,usenatbib]{mnras}
\usepackage[percent]{overpic}
\usepackage{times}
\usepackage{graphicx}
\usepackage{amsmath}
\usepackage{mathabx}
\usepackage{subfigure}
\usepackage{natbib}
\usepackage{txfonts}
\usepackage{journals}
\usepackage{xcolor}
\usepackage[utf8]{inputenc}
\usepackage[overload]{empheq}
\usepackage{verbatim}

\newcommand{\Ha}{${\mathrm{H\alpha}}$}

\newcommand{\gb}{$G_{\rm BP}$}
\newcommand{\gr}{$G_{\rm RP}$}

\newcommand{\ebr}{$E(G_{\rm BP}-G_{\rm RP})$}
\newcommand{\br}{$(G_{\rm BP}-G_{\rm RP})$}
\newcommand{\rbr}{$R_{(G_{\rm BP}-G_{\rm RP})}$}
\newcommand{\bri}{$(G_{\rm BP}-G_{\rm RP})_0$}

\newcommand{\ebv}{$E(B-V)$}
\newcommand{\arc}{$^{\prime\prime}$}
\newcommand{\ar}{$A_{r}$}

\title[Reddening of LAMOST M dwarfs]
{Estimating accurate reddening values of LAMOST M dwarfs} 

\author[H.~Shen et al.]
{H. Shen,$^{1}$
B.-Q. Chen,$^{1}$\thanks{E-mail: bchen@ynu.edu.cn.}
H.-L. Guo,$^1$
H.-B. Yuan,$^{2}$
W.-X. Sun,$^1$
and  J. Li$^3$
\\
$^{1}$South-Western Institute for Astronomy Research, Yunnan University, Kunming, 650500, P.\,R.\,China\\
$^{2}$Department of Astronomy, Beijing Normal University, Beijing, 100875, P.\,R.\,China\\
$^{3}$Physics and Astronomy College, China West Normal University, Nanchong 637002, P.\,R.\,China\\
}
\begin{document}

\date{Accepted ???. Received ???; in original form ???}

\pagerange{\pageref{firstpage}--\pageref{lastpage}} \pubyear{2022}
\maketitle
\label{firstpage}

\begin{abstract}
M dwarfs are the dominating type of stars in the solar neighbourhood. They serve as excellent tracers for the study of the distribution and properties of the nearby interstellar dust. In this work, we aim to obtain high accuracy reddening values of M dwarf stars from the Large Sky Area Multi-Object Fiber Spectroscopic Telescope (LAMOST) Data Release 8 (DR8). Combining the LAMOST spectra with the high-quality optical photometry from the Gaia Early Data Release 3 (Gaia EDR3), we have estimated the reddening values \ebr\ of 641,426 M dwarfs with the machine-learning algorithm Random Forest regression. The typical reddening uncertainty is only 0.03\,mag in \ebr. We have obtained the reddening coefficient \rbr, which is a function of the stellar intrinsic colour \bri\ and reddening value \ebv. The values of \ebv\ are also provided for the individual stars in our catalogue. Our resultant high accuracy reddening values of M dwarfs, combined with the Gaia parallaxes, will be very powerful to map the fine structures of the dust in the solar neighbourhood.     
\end{abstract}


\section{Introduction} \label{sec:intro}

Our Sun locates in the Local Bubble, which suffers from less dust extinction compared to other places in the Galactic disk \citep{Lallement2014, Chen2014, Chen2019, Green2019, Leike2020}. To explore the properties and fine structures of the dust nearby, high accuracy reddening values of the individual local stars are always welcome.

M dwarfs are faint stars with low mass and low surface temperature \citep{Laughlin1997, Pecant2013}. They are the most common type of stars in the solar neighbourhood \citep{Bochanski2010, Winters2019}. They can thus be utilised to study the dust properties and structures in the solar neighbourhood. \cite{Jones2011} used the Sloan Digital Sky Survey (SDSS) spectra of over 56,000 M dwarfs to map the dust distribution in the high Galactic latitudes of the local Galaxy. They compare the SDSS spectra of the target M dwarfs to a set of template M dwarf spectra from low extinction regions to derive the extinction values of the individual stars. Their resulted extinction values of the M dwarfs have a median value of 0.037\,mag in $A_V$ and an internal uncertainty of about 0.043\,mag in $A_V$. However, the size of the \citet{Jones2011} M dwarf catalogue is relatively small and most of their sources are located in high Galactic latitudes. 

The ongoing Large Sky Area Multi-Object Fiber Spectroscopic Telescope (LAMOST; \citealt{Cui2012}) Galactic survey \citep{Deng2012, Zhao2012, Liu2014} have obtained more than ten million spectra of stars in the Milky Way. \citet{Zhong2015b,Zhong2015a} and \citet{Li2016} have developed a template-fitting pipeline to identify and classify M stars from the LAMOST spectra. \citet{Zhong_2019} have provided a catalogue of 39,796 M giants and 501,152 M dwarfs based on the LAMOST Data Release 5 (DR5). \citet{Li2021} and \citet{Du2021} have developed new pipelines to calculate the atmospheric parameters, including the effective temperature $T_{\rm eff}$, metallicity [M/H] and surface gravity log\,$g$, of the LAMOST M stars. The most recently released catalogues from the LAMOST Data Release 8 (DR8; \citealt{Luo2015}) contains an M type star catalogue which consists of 721,947 M dwarf spectra, 44,312 M giant spectra and 4,316 M sub-dwarf spectra. In this work, we will combine the LAMSOT M dwarf star catalogue with the Gaia Early Data Release 3 (Gaia EDR3; \citealt{Gaiaedr3-2021}) photometry to estimate the reddening values of the individual M dwarfs. 

\section{Data}

\begin{figure*}
\centering
\includegraphics[width=0.59\textwidth]{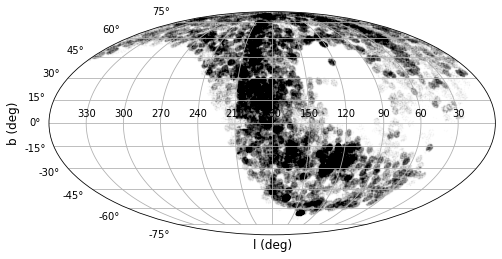}
\includegraphics[width=0.39\textwidth]{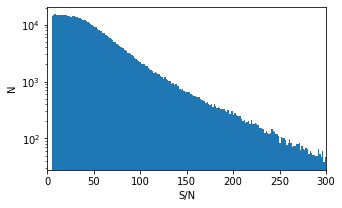}
\caption{Spatial distribution (upper panel) and the histogram of the LAMOST $i$-band S/N (bottom panel) of the 648,220 selected M dwarf stars.}  
\label{spadist}
\end{figure*}

Our work is based on the LAMOST DR8 v1.1\footnote{\url{http://www.lamost.org/dr8/v1.1/}}, which was recently published in February 2022. LAMOST DR8 v1.1 contains both the low and medium resolution spectra and catalogues obtained by the LAMOST between October 2011 and May 2020. In the current work, we adopt the M star catalogue named `LAMOST LRS Catalog of gM, dM, and sdM stars'. The catalogued sources are selected from the low resolution survey. The LAMOST low resolution spectra have a resolution of R $\sim$ 1800 at 5500\,\AA\ and cover the full optical wavelength ranging between 3800 and 9000\,\AA. The LAMOST M star catalogue contains 44,312 M giant spectra, 721,947 M dwarf spectra, and 4,316 M sub-dwarf spectra. The catalogue present values of atmospheric parameters ($T_{\rm eff}$, log\,$g$ and [M/H]; \citealt{Du2021}), equivalent widths (EW) of the \Ha, TiO, CaH, CaOH and Na lines, and a metallicity sensitive parameter of ``zeta'' deduced from the LAMOST spectra. The parameter ``zeta'' is defined as $\zeta_{\rm TiO/CaH} $ by the Eq.~(2) of \citet{Lepine2007}. It can be calculated from the CaH2, CaH3 and TiO5 indices obtained from the stellar spectrum.  By definition, stars with metallicities equal to the Sun have $\zeta_{\rm TiO/CaH} =1$, and stars with metallicities larger and smaller than the Sun have $\zeta_{\rm TiO/CaH} $  smaller and larger than 1, respectively. 

To obtain the precise reddening values of the LAMOST M dwarfs, we adopt the high quality photometry from Gaia EDR3, which was released in December 2020 \citep{Gaiaedr3-2021}. Gaia EDR3 contains about 1.8 billion sources with high precision celestial positions and $G$ broad-band photometric measurements. Among them, $\sim$ 1.5 billion sources have \gb- and \gr-band magnitudes and five astrometric parameters (position, parallaxes and proper motions). The typical uncertainties for the Gaia EDR3 $G$, \gb\ and \gr\ photometries are respectively 6, 108 and 52\,mmag at G = 20 mag \citep{Gaiaedr3-2021}.

We select the M dwarfs from the LAMOST M star catalogue and cross-match them with the Gaia EDR3 using the CDS XMatch Service. We adopt a matching radius of 3\arc\ which is the diameter of the LAMOST low resolution fibre. Most (81\,per\,cent) of our catalogued stars have total proper motions smaller than 50\,mas\,yr$^{-1}$ measured by Gaia. There are only 33 stars having proper motions larger than 200\,mas\,yr$^{-1}$. The coordinates of LAMOST stars are adopted from the photometric data such as the multiband CCD photometric survey of the Galactic Anticentre with the Xuyi 1.04/1.20\,m Schmidt Telescope (XSTPS-GAC; observed during 2009 and 2011; \citealt{Liu2014}) and the Pan-STARRS 1 Survey (PS1; observed during 2010 and 2014; \citealt{Chambers2016}). The Gaia EDR3 are from the observations of Gaia during 2014 and 2017. If we assume a typical 5\,yr epoch difference between the LAMOST and Gaia coordinates, the moving distances for most of our catalogued stars are smaller than $\sim$ 0.25\,arcsec. Only 33 stars have moving distances larger than 1\,arcsec. Therefore, the mismatches of stars caused by the stellar proper motions can be ignored in our work. We select sources with LAMOST spectra having signal-to-noise ratios (S/N) in $i$-band (snri) larger than 5, having EW measurements for all the \Ha, TiO, CaH, CaOH and Na lines and having valid zeta values. For the Gaia photometry, we select stars having both the \gb- and \gr-band magnitudes and exclude sources with bad Gaia photometry by phot\_bp\_rp\_excess\_factor $<$ $1.3 + 0.06(G_{\rm BP}-G_{\rm RP})^2$. These requirements lead to 648,220 sources in our catalogue. The spatial distribution and the histogram of the $i$-band S/N of all the sample stars are shown in Fig.~\ref{spadist}.

\section{Reddening determinations}

\begin{figure}
\centering
\includegraphics[width=0.45\textwidth]{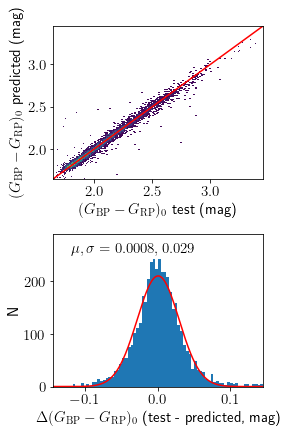}
\caption{Comparison of intrinsic colours \bri\ as predicted by the Random Forest model and those calculated from the Planck reddening values for the test sample stars. The red straight line in the upper panel denotes complete equality to guide the eyes. In the bottom panel, the red curve is a Gaussian fit to the histogram of the differences. The mean and dispersion of the differences are marked in the diagram.}  
\label{traincomp}
\end{figure}

To obtain the reddening values of our catalogued M dwarfs, we have built a Random Forest regression model that returns the intrinsic colours \bri\ of stars. The Random Forest model is one of the most effective machine-learning models for regression problems. It fits a multitude of decision trees to the individual sub-samples of the data set and adopts the average predicted values to control over-fitting and improve accuracy.

To train the Random Forest model, we define a training sample of high quality measurements and suffering from very small dust extinction effects by imposing the following criteria:
\begin{enumerate}
 \item LAMOST spectra S/N in the $i$-band snri $>$ 30 and photometric uncertainties of both the Gaia EDR3 \gb\ and \gr\ magnitudes $<$ 0.01\,mag,
 \item reddening values \ebv\ from the Planck two-dimensional dust maps \citep{Planck2014} $<$ 0.02\,mag,
 \item matching distances between the LAMOST and Gaia positions $<$ 1\arc.
 \end{enumerate}
This leads to a sample of 21,150 stars. The intrinsic colour \bri\ of these sample stars are calculated using the reddening values from \citet{Planck2014} dust maps and the reddening coefficients from \citet{Chen2019}, by $(G_{\rm BP} - G_{\rm RP})_0 = (G_{\rm BP} - G_{\rm RP}) - 1.33E(B-V)$. 

We randomly select 80\,per\,cent stars from the training sample to train our Random Forest model and use the remaining 20\,per\,cent stars to test the training accuracy. In the current work, we set the equivalent widths (EW) of the \Ha, TiO, CaH, CaOH and Na lines (ewha, tio1, tio2, tio3, tio4, tio5, cah1, cah2, cah3, caoh and na) and the metallicity sensitive parameter (zeta) deduced from the LAMOST spectra as the input parameters. The {\sc python} package {\sc scikit-learn} \citep{Pedregosa2011} is adopted to build the model. We tried different parameters with grid search to optimize the models. Finally, we adopt the number of features to consider when looking for the best split \verb|max_features = `sqrt'|, the minimum number of samples required to be at a leaf node \verb| min_samples_leaf = 4|, the minimum number of samples required to split an internal node \verb|min_samples_split = 5| and the number of trees in the forest \verb| n_estimators = 581|. The best model gets a $R^2$ regression score r2\_score = 0.95. We compare the values of \bri\ predicted by the Random Forest model and those calculated from the reddening values from the Planck dust map for the test stars in Fig.~\ref{traincomp}. The dispersion of the differences is only $\sim$ 0.03\,mag in \ebr.\ Finally, the trained Random Forest model is applied to all the 648,220 sources to predict their intrinsic colours \bri. The reddening values of these sources are then obtained by the standard relation $E(G_{\rm BP} - G_{\rm RP}) = (G_{\rm BP} - G_{\rm RP}) - (G_{\rm BP} - G_{\rm RP})_0$.  

\begin{table*}
    \centering
    \begin{tabular}{lll}
    \hline
    Column  & Name   & Description \\
    \hline 
     1 & spec\_id & Unique spectral ID of LAMOST sources, in format of `date-plateid-spectrographid-fibreid' \\
     2 & sourceid & Gaia EDR3 sourceid \\
     3 & ra & Right ascension (\degr) \\
     4 & dec & Declination (\degr) \\
     5 & snr & LAMOST S/N in $i$-band \\
     6 & bp & Gaia EDR3 \gb\ magnitude \\
     7 & bp\_err & Gaia EDR3 \gb\ magnitude uncertainty \\
     8 & rp & Gaia EDR3 \gr\ magnitude \\
     9 & rp\_err & Gaia EDR3 \gr\ magnitude uncertainty \\     
     10 & ebprp & Reddening \ebr\ values resulted in the current work \\
     11 & ebprp\_err & Error of \ebr \\
     12 & ebv & Reddening \ebv\ values, converted from \ebr \\
    \hline
    \end{tabular}
    \caption{Description of the M dwarfs reddening catalogue.}
    \label{tab1}
\end{table*}

\begin{figure}
\centering
\includegraphics[width=0.49\textwidth]{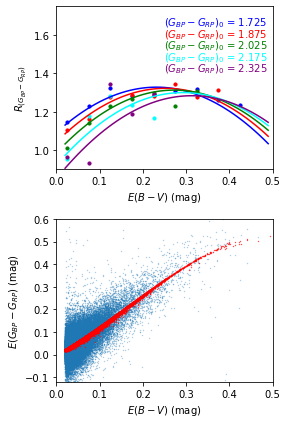}
\caption{Fitting results of the reddening coefficient \rbr. In the upper panel, the dots and curves are respectively the measured and fitted \rbr\ values. Different colours denote different stellar intrinsic colours, which are labeled in the diagram. In the bottom panel we show reddening \ebr\ versus \ebv. The blue and red dots are respectively the measured and fitted values.}  
\label{rbr}
\end{figure}

\begin{figure}
\centering
\includegraphics[width=0.49\textwidth]{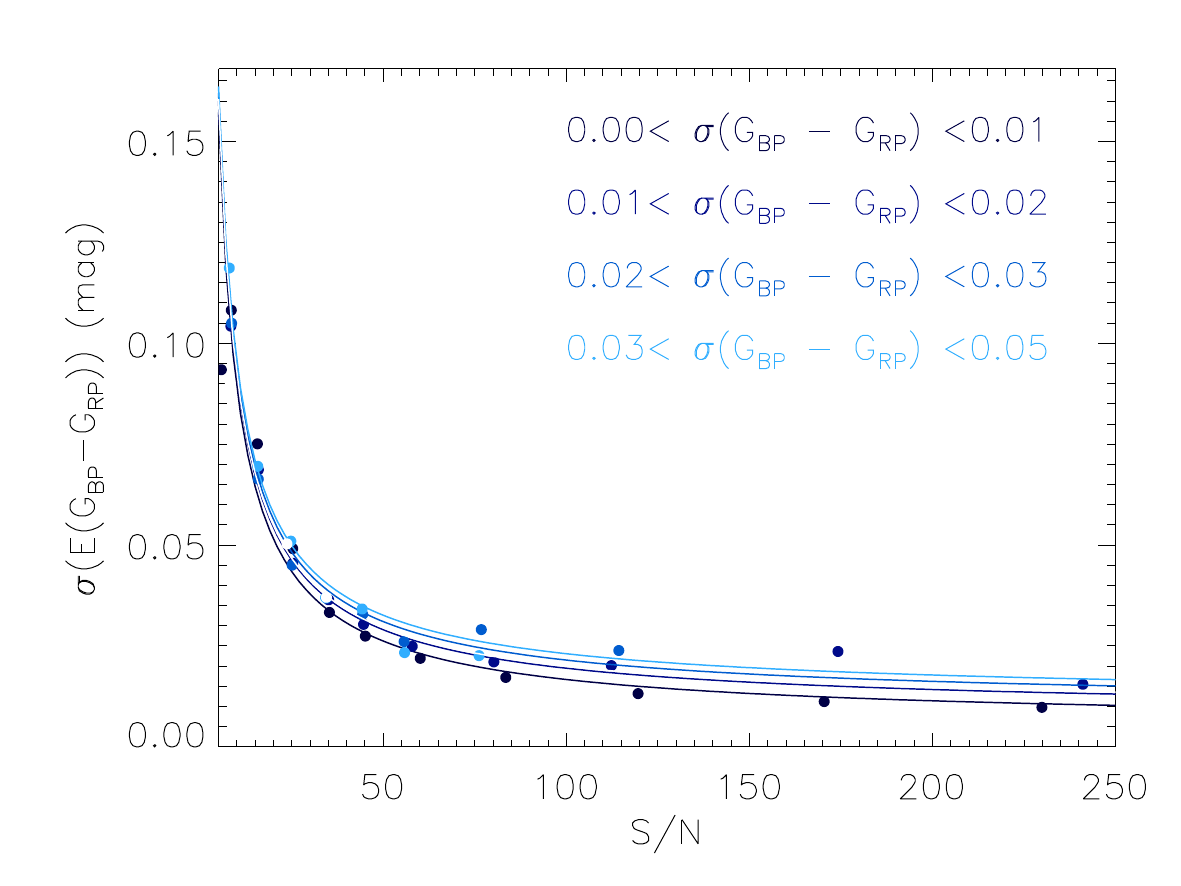}
\caption{Random uncertainties of reddening $\sigma$(\ebr), estimated by comparing results of duplicate observations of similar S/N values, as a function of the S/N and $\sigma$\br\ values. The dots and curves are respectively the measured and fitted values. Different colours denote different $\sigma$\br\ bins, which are labeled in the diagram.}  
\label{rnderr}
\end{figure}

\begin{figure*}    	
\centering    	
\includegraphics[height=5.7cm]{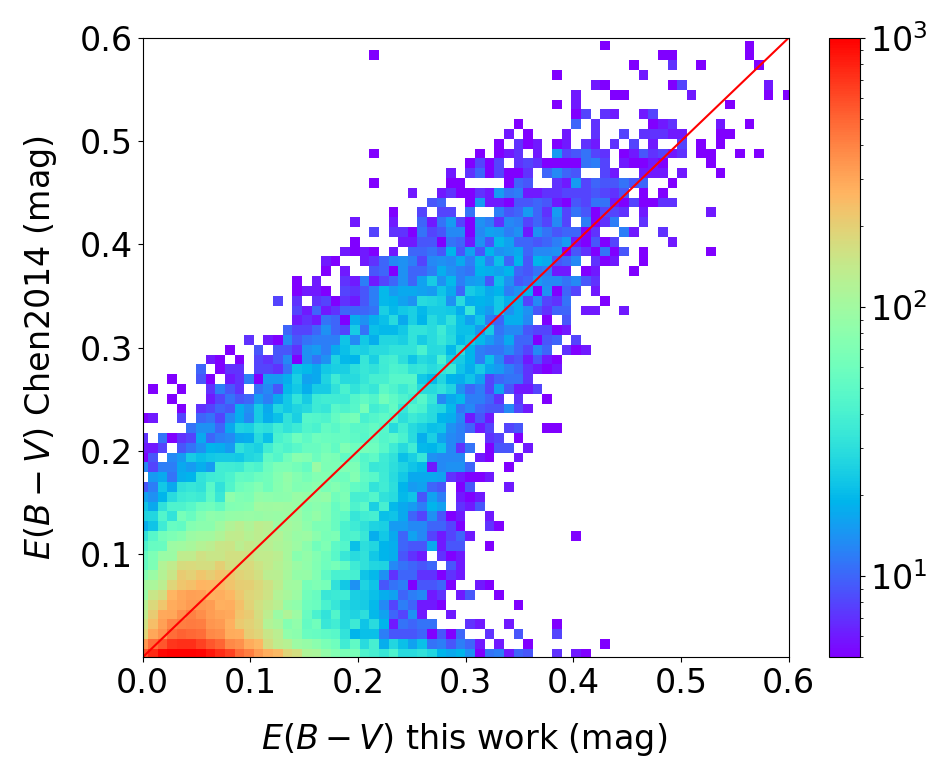}    		 
\includegraphics[height=5.7cm]{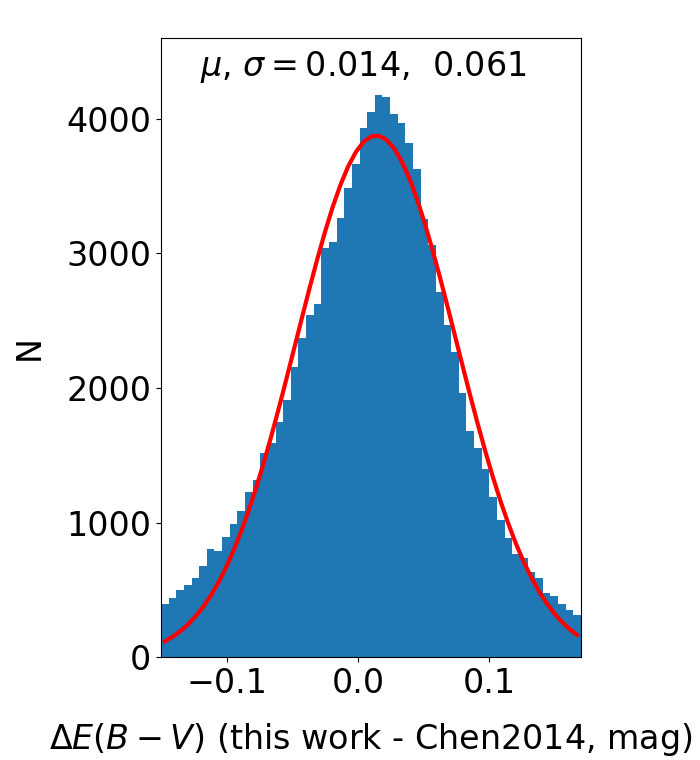}    \\		 
\includegraphics[height=5.7cm]{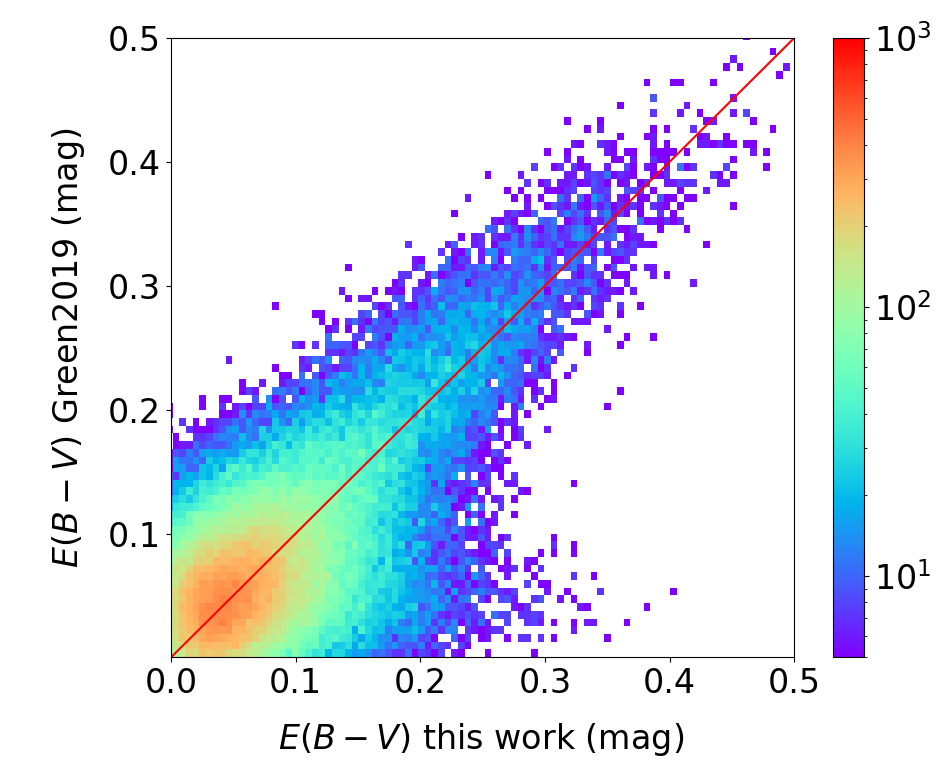}       	 	
\includegraphics[height=5.7cm]{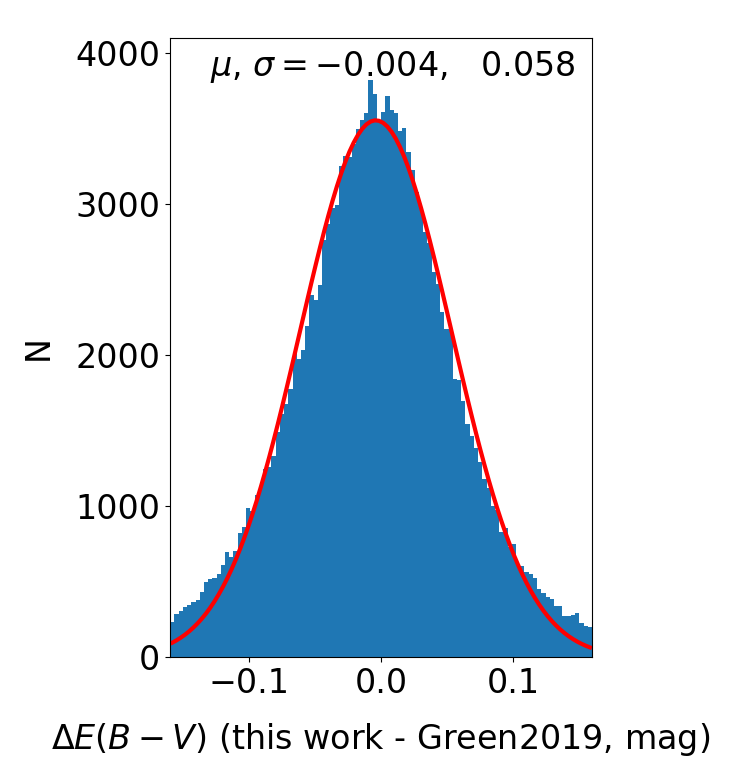}   \\     	
\includegraphics[height=5.7cm]{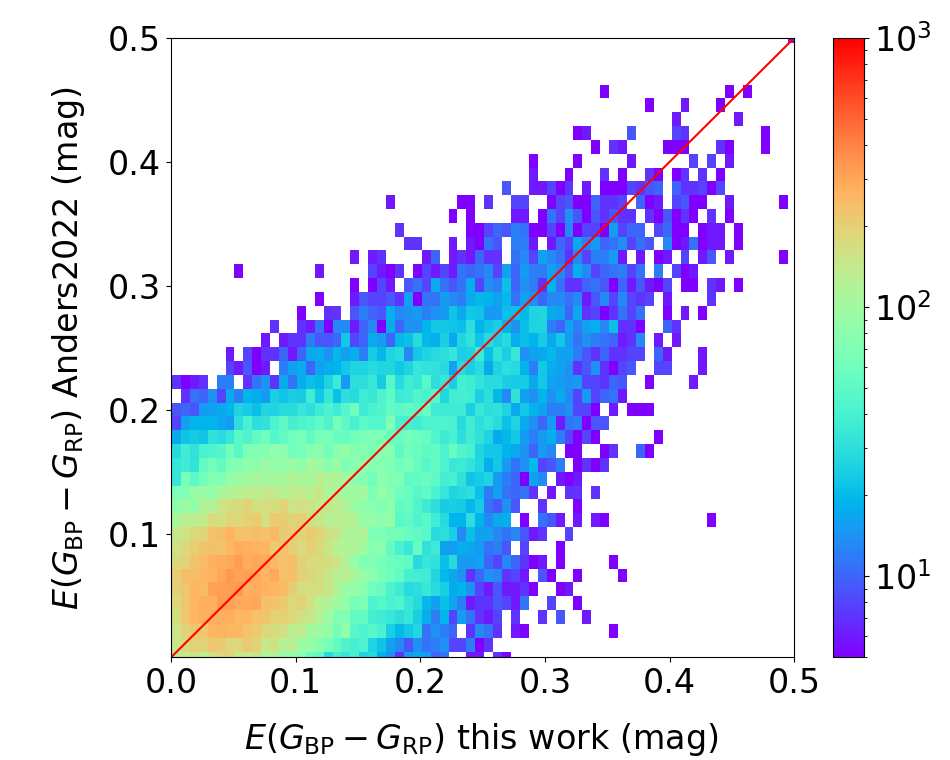}    
\includegraphics[height=5.7cm]{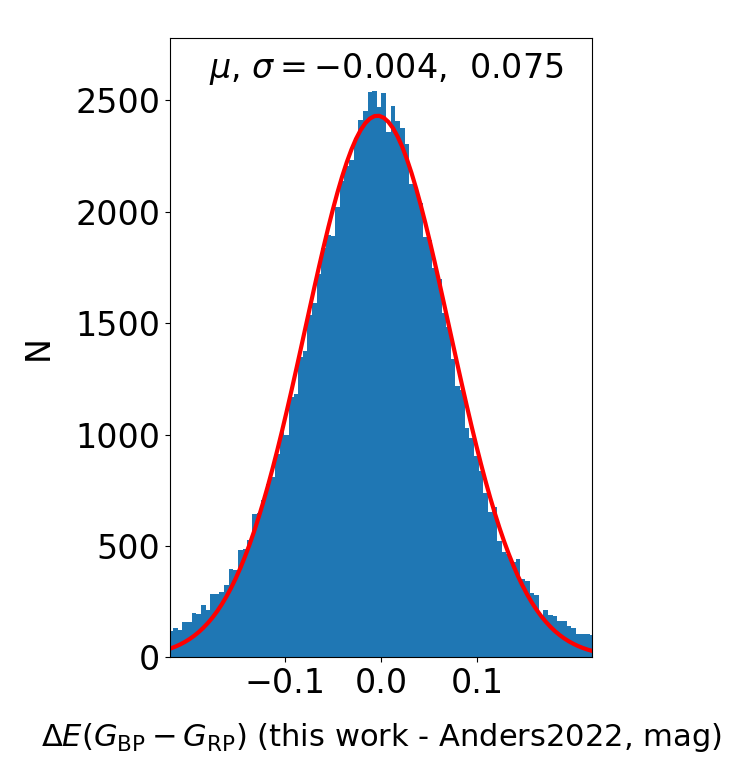}        	
 \caption{Comparisons of the reddening values in our work and those from \citet[top panels]{Chen2014}, \citet[middle panels]{Green2019} and \citet[bottom panels]{Anders2022}. The red straight lines in the left panels denoting the complete equality are over-plotted to guide the eyes. Red curves in the right panels are Gaussian fits of the distributions of the reddening differences. The means and standard deviations of the Gaussian fits are labeled in the diagrams.}  
 \label{comparison_bp_rp}
 \end{figure*}

 \begin{figure}
\centering
\includegraphics[width=0.49\textwidth]{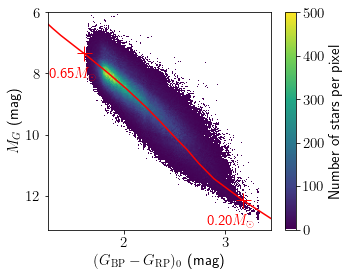}
\caption{Dust-free colour and absolute magnitude diagram of the sample M dwarfs with Gaia EDR3 parallax error less than 20\,per\,cent. The red curve represent a PARSEC isochrone with the Solar metallicity and age of 10\,Gyr \citep{PAdova2012}. The two red ``+'' symbols mark the positions of stars with initial masses of 0.2 and 0.65\,$M_{\odot}$.}  
\label{cmd}
\end{figure}
 
 \begin{figure*}
\centering
\includegraphics[width=0.59\textwidth]{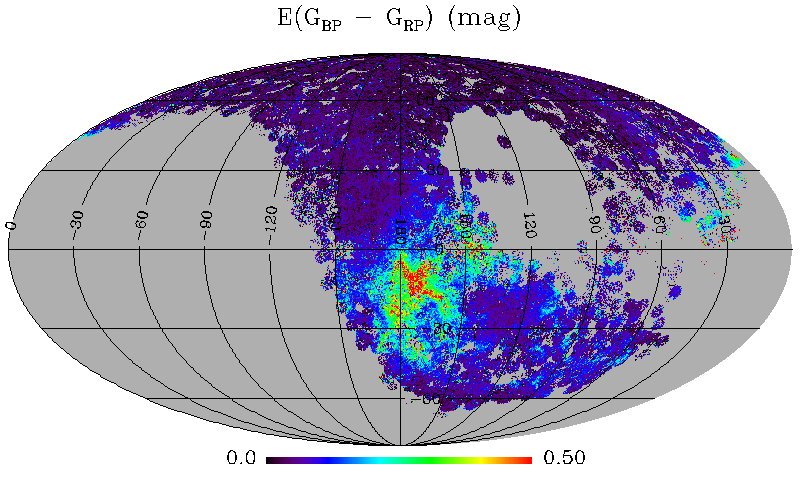}
\caption{Map of our \ebr\ values in the Galactic longitude $l$ and latitude $b$ based on all our catalogued stars. The map has a resolution of 27\,arcmin.}  
\label{lbdist}
\end{figure*}

\begin{figure*}
\centering
\includegraphics[height=5.67cm]{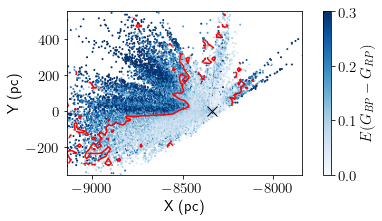}
\includegraphics[height=5.67cm]{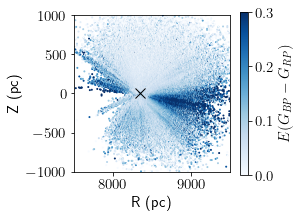}
\caption{{\it Left panel:} Map of our \ebr\ values in the $X$-$Y$ plane for stars with $|Z| < 50$\,pc (averaged over 5$\times$5\,pc$^2$). The red contours of \ebr\ = 0.15\,mag is overplotted in the diagram. {\it Right panel:} Map of \ebr\ values in the $R$-$Z$ plane for all the catalogued stars (averaged over 5$\times$5\,pc$^2$). The ``x'' symbols in both the panels mark the position of the Sun ($X$ = $-$8.34\,kpc, $Y$ = 0\,kpc, $R$ = 8.34\,kpc and $Z$ = 0\,kpc).}  
\label{xyrz}
\end{figure*}

\section{Results and discussion}

We have obtained reddening values for all the catalogued LAMOST M dwarfs. As most of the M dwarfs are located nearby, they usually suffer from relatively small dust reddening effects. The intrinsic colours \bri\ of the sources predicted by our Random Forest model could be larger than the observed values due to uncertainties of the observations and our method, which would lead to negative values. In the current work, we keep only the sources with resulted reddening \ebr\ values larger than $-$0.25\,mag. This leads to a final sample of 641,426 stars. The catalogue of M dwarfs with resulting reddening values is available online via ``\url{http://paperdata.china-vo.org/diskec/lstmdwarfs/table1.fits}''. The data format of the catalogue is described in Table~\ref{tab1}. 

\subsection{Converting $E(G_{\rm BP}-G_{\rm RP})$ to $E(B-V)$}

In addition to the \ebr\ values, we also provide the mostly used \ebv\ reddening values for the individual stars in our catalogue. In the current work, we have adopted a reddening coefficient of \rbr $=$ 1.33 from \citet{Chen2019} to convert the Planck \ebv\ to \ebr\ for the stars in the training data set. The reddening coefficient \rbr\ should vary for the individual stars, due to: 1) the variation of the extinction curves in different environments \citep{Jones2011, Schlafly2016}, 2) the wide wavelength ranges of the Gaia \gb\ and \gr\ bands, and 3) the variation of the stellar spectral energy distributions (SEDs). Stars in the training sample are all located in very low extinction regions, with line-of-sight reddening \ebv $<$ 0.02\,mag. The variation of the reddening coefficient \rbr\ would not affect our results. However, to convert \ebr\ to \ebv\ for all our catalogued stars accurately, we should consider the variation of the \rbr.

Physically, \rbr\ is related to the stellar intrinsic colour \bri\ and the reddening \ebv\ \citep{Niu2021_2,Niu2021_1, Sun2022}. To explore the correlation between \rbr\ and (\bri, \ebv), we have selected stars with large distances to the Galactic plane ($Z$), where the stars are thought to be out of the Galactic dust disk and we assume their \ebv\ values from the \citet{Planck2014} dust map are reliable. We adopt the following criteria:
\begin{enumerate}
 \item LAMOST spectra S/N in the $i$-band snri $>$ 30 and photometric uncertainties of both the Gaia EDR3 \gb\ and \gr\ magnitudes $<$ 0.01\,mag,
 \item reddening values \ebv\ from the Planck two-dimensional dust maps \citep{Planck2014} between 0.02 and 0.5\,mag,
 \item Galactic latitude $|b|$ $>$ 15\degr,
 \item Distance to the Galactic plane $|Z|$ $>$ 300\,pc,
 \item matching distances between the LAMOST and Gaia positions $<$ 1\arc.
 \end{enumerate}
This leads to a sample of 62,905 stars. We note that stars with very red colours (\br $>$ 2.5\,mag) are not included in this selected sample, since they are too faint to be observed at $|Z| > $ 300\,pc. The reddening coefficients \rbr\ of the selected stars are calculated by \rbr~$=$~\ebr$/$\ebv, where \ebr\ are calculated in the current work and \ebv\ are adopted from the \citet{Planck2014} dust map. The selected stars are divided into 10 $\times$ 10 pixels in the \bri\ and \ebv\ space. We obtain the averaged \rbr\ value of each pixel and fit \rbr\ as a function of \bri\ and \ebv. As a result, we obtain,
\begin{equation}
\begin{aligned}
    R_{(G_{\rm BP} - G_{\rm RP})} = & 1.148 -0.014E(B-V) -4.418E(B-V)^2  \\ 
    & +0.239(G_{\rm BP} - G_{\rm RP})_0 -0.158(G_{\rm BP} - G_{\rm RP})_0^2  \\ 
    & +1.195E(B-V)(G_{\rm BP} - G_{\rm RP})_0. 
\end{aligned}
\end{equation}
In Fig.~\ref{rbr} we show the fitting results. Our derived correlations between \rbr\ and (\bri, \ebv) are very similar as those derived from \citet[see their Fig.~1]{Sun2022}, which increase our confidence that both relations are robust. We have converted our resulted \ebr\ values to \ebv\ for all the stars. For stars with colours redder than 2.325\,mag, we adopt the reddening coefficients calculated with the colour \bri = 2.325\,mag. The derived \ebv\ values are also listed in our catalogue.

\subsection{Reddening uncertainties}

For the uncertainties of our derived reddening \ebr, we first consider the random errors, which are estimated from the dispersions of reddening differences (divided by $\sqrt{2}$) yielded by duplicate observations. We have found 96,118 unique stars in our catalogue that have more than one spectra. We select stars with duplicate spectra that have similar $i$-band S/N and compare their resulted reddening values. Fig.~\ref{rnderr} shows the difference dispersions divided by $\sqrt{2}$, i.e. the random errors, as a function of the $i$-band S/N of the LAMOST spectra and the photometric errors of the Gaia \br\ colours $\sigma$\br\ computed by adding the photometric errors in quadrature. For stars of LAMOST S/N of 10, the random errors are about 0.10\,mag. The random errors decrease to $\sim$ 0.03\,mag for stars with S/N of about 50 and becomes better than $\sim$ 0.02\,mag at S/N values larger than 100. We fit the random errors as a function of the S/N and $\sigma$\br\ values. As a result, we obtain,
\begin{equation}
\begin{aligned}
    \sigma(E(G_{\rm BP} - G_{\rm RP})) = & 0.003 + 0.357\sigma(G_{\rm BP} - G_{\rm RP}) \\
    & -3.882\sigma(G_{\rm BP} - G_{\rm RP})^2 + \dfrac{0.602}{{\rm S/N}^{0.854}},
\end{aligned}    
\end{equation}
which are used to calculate the random errors of the individual stars in our catalogue. For stars of S/N better than 200, their random errors are assigned to be values of those at S/N $=$ 200. The typical random error of our catalogued stars is about 0.03\,mag in \ebr.  

In addition to the random errors described above, the uncertainties of our derived reddening values also contain systematic errors which arise from the fact that our training data set are not from perfectly extinction-free lines of sight. Our training sample stars were chosen from regions of integrated reddening of up to 0.02\,mag in \ebv. The median line-of-sight reddening of our training sample stars is \ebv $=$ 0.016\,mag. Since most of the training sample stars are located at high Galactic latitudes ($b > 15$\degr) and large distances to the Galactic plane ($|Z| > 300$\,pc), the deviation between their actual reddening values and those from the \citet{Planck2014} dust map should be negligible. Because of this, we have ignored the systematical uncertainties in the current work and adopted only the random errors as the final reddening uncertainties. An alternative method is to obtain the intrinsic reddening values of the training sample star from a synthetic spectral library. However, it would result in additional errors from the uncertainties of the synthetic stellar models \citep{Yuan2015}. 

\subsection{Comparison with previous work}

The typical error of our resultant reddening values is estimated to be $\sim$ 0.03\,mag. To verify the robustness of our reddening values, we have compared our results with those from previous works. The comparisons are shown in Fig.~\ref{comparison_bp_rp}. \cite{Chen2014} obtained $r$-band extinction values for over 13 million stars in the Galactic anticentre from the SED fits to the multi-band photometric measurements of the Xuyi Schmidt Telescope Photometric Survey of the Galactic Anticentre (XSTPS-GAC), the Two Micron All Sky Survey (2MASS) and Wide-Field Infrared Survey Explorer (WISE). We cross-matched our catalogue with that from \citet{Chen2014} with a radius of 3\arc, which yields 112,621 common stars. The Chen et al. extinction values are converted to reddening values by the extinction coefficients from \cite{Yuan2013}: \ebv = 0.43\ar. Our resulted reddening values are in good agreement with those from \citet{Chen2014}. The mean difference is 0.014\,mag and the dispersion is 0.061\,mag.

Based on Gaia parallaxes and stellar photometry from Pan-STARRS\,1 (PS1) and 2MASS, \cite{Green2019} obtained reddening values of 799 million stars in the northern sky. To compare the Green et al. results with our values, we have selected stars in the Green et al. catalogue by the following criteria: 1) stars been detected in all the eight PS1 and 2MASS  passbands ($g_{\rm P1}$, $r_{\rm P1}$, $i_{\rm P1}$, $z_{\rm P1}$, $y_{\rm P1}$, $J$, $H$ and $K_{\rm S}$) and the photometric errors in all the eight bands smaller than 0.05\,mag; 2) the goodness-of-fit (maximum-likelihood $\chi^2/$passband) smaller than 5; and 3) the uncertainties of reddening smaller than  0.08\,mag. The selected stars are cross-matched with our catalogue and we have obtained 173,603 common stars. Our results agree well with those selected from the Green et al. catalogue, with a negligible averaged difference and a small dispersion of only 0.058\,mag. 

\citet{Anders2022} presented a catalogue of 362 million stars. Their reddening values are derived from the photometric catalogues of Gaia EDR3, PS1, SkyMapper, 2MASS, and WISE. We select stars from the Anders et al. catalogue by the following criteria: 1) stars with good astrometric measurements (astrometric fidelity flag larger than 0.5; \citealt{Rybizki2022}) and good \gb\gr\ photometries ($\lvert C_* \rvert / {\sigma_{C^*}} < 0.5$, where $C_*$ and ${\sigma_{C^*}}$ are respectively the bp\_rp\_excess\_corr value and a simple function of the Gaia $G$ magnitude defined in \citealt{Riello2021}) ; 2) stars been detected in all the Gaia $G$\gb\gr, the 2MASS $JHK_{\rm S}$, the WISE $W1W2$ and the PS1/SkyMapper $griz$ bands; 3) stars with good fitting models (the first number of the flag `sh\_outflag' $=$ 0) and small fitting errors (the third number of `sh\_outflag' $=$ 0). The selected stars are then cross-matched with our catalogue, which yields 99,273 common stars. The agreement of the comparison between our results and those from \citet{Anders2022} is good. The average difference is negligible and the dispersion is 0.075\,mag in \ebr.

\subsection{The Hertzsprung--Russell diagram}

Based on the stars for which we measured accurate reddening values, we have plotted the Hertzsprung–Russell diagram (HRD) of the LAMOST M dwarfs (Fig.~\ref{cmd}). We have selected 528,368 stars in our catalogue that have Gaia EDR3 parallax errors smaller than 20\,per\,cent. The distances of the selected stars $d$ are calculated from the Gaia EDR3 parallaxes $\varpi$ by a simple Bayesian algorithm \citep{BailerJones2018, Chen2019OBstar}. We adopt a posterior probability given by,
\begin{equation}
    p(d|\varpi) = d^2\exp(-\dfrac{1}{2\sigma^2_{\varpi}}(\varpi-\varpi_{\rm zp} - \dfrac{1}{d}))p(d),
\end{equation}
where $\sigma_{\varpi}$ is the Gaia parallaxes uncertainties, $\varpi _ {\rm zp}$ the global parallax zero point and $p(d)$ the space density distribution prior for the M dwarfs.
In the current work, we adopt $\varpi _ {\rm zp}$ $=$ $-$0.026\,mas from \citet{Huang2021} and the Galactic disk model derived from the late type stars (1.0 $<~r-i~<$ 1.4\,mag) by \citet{Juric2008}. 

With the resultant distances, the Gaia $G$-band absolute magnitudes are then calculated by $M_G = G - 5\lg\,d + 5 - A_G$. In the current work, we adopted the $G$-band extinction coefficient from \citet{Chen2019} to obtain the $G$-band extinction, by $A_G = 1.91E(B-V)$. We show the dust-corrected colour and absolute magnitude diagram for the selected 528,368 stars in Fig.~\ref{cmd}. A PARSEC isochrone with the Solar metallicity and age of 10\,Gyr \citep{PAdova2012} is overplotted in the diagram. The distribution of our catalogued stars is in good agreement with the theoretic isochrone, which suggests the robustness of our resulted reddening values. Most of our catalogued stars have intrinsic colours \bri\ between 1.6 and 3.2\,mag and absolute magnitudes $M_G$ between 7 and 12\,mag. According to the PARSEC isochrone, our catalogued stars have masses between 0.2 and 0.65\,$M_{\odot}$.

\subsection{The Reddening distribution}

Finally, we plot the \ebr\  reddening maps in the Galactic longitude $l$ and latitude $b$ coordinates and the Galactocentric coordinates based on our catalogued M dwarfs. In Fig.~\ref{lbdist}, we show the reddening map in the Galactic longitude $l$ and latitude $b$. We use the HEALPix pixelization scheme \citep{Gorski2005} to bin the sample stars into individual pixels. We adopt a resolution of 27\,arcmin (with HEALPix nside = 128) and average the reddening values of stars in the individual pixels. Fig.~\ref{lbdist} presents a nice local dust reddening map, which show very similar features to the nearest distance slice of Galactic three-dimensional (3D) dust maps, such as the \citet{Green2019} 3D extinction map with distance 0$~<~d~<~$0.5\,kpc (their Fig.~1) and the \citet{Guo2021} map with 0$~<~d~<~$0.4\,kpc (their Fig.~14). The famous nearby giant molecular clouds, such as the Orion ($l,~b$ $\sim$ $-$160\degr, $-$15\degr), the Taurus ($l,~b$ $\sim$ 170\degr, $-$15\degr), the Perseus ($l,~b$ $\sim$ 155\degr, $-$20\degr), the Aries ($l,~b$ $\sim$ 160\degr, $-$30\degr), the Eridanus ($l,~b$ $\sim$ 190\degr, $-$35\degr), the Aquila Rift ($l,~b$ $\sim$ 30\degr, 10\degr) and the Pegasus ($l,~b$ $\sim$ 90\degr, $-$35\degr), are clearly visible in the map.

We have also created averaged \ebr\ maps in the Galactocentric coordinates based on the 528,368 stars with Gaia EDR3 parallax errors smaller than 20\,per\,cent. The maps are presented in Fig.~\ref{xyrz}. In the left panel of Fig.~\ref{xyrz}, we show the distribution of reddening \ebr\ values in the Galactic $X$-$Y$ plane for stars with distances to the Galactic plane within 50\,pc. The most significant feature in the Figure is the Local Bubble, which are marked by the \ebr = 0.15\,mag contour lines. The shape of the Local Bubble is in good agreement with those derived from the previous 3D extinction maps (e.g. Fig.~12 from \citealt{Chen2019} and Fig.~4 from \citealt{Lallement2022}). In the right panel of Fig.~\ref{xyrz} we show the averaged reddening \ebr\ maps in the Galactic cylindrical coordinates $R$ and $Z$. The Figure displays a remarkable shape of the Galactic dust disc. Based on our catalogued M dwarfs with accurate reddening values, we will be able to investigate the large and small scales dust structure in the solar neighbourhood. More quantitative analysis will be presented in future works.

\section{Conclusions}

By combining the LAMOST DR8 M dwarf star catalogue and Gaia EDR3, we have presented a catalogue containing accurate reddening values \ebr\ of 641,426 M dwarfs, which are estimated by the Random Forest regression. With the benefit of the LAMOST spectral information and the Gaia accurate photometric observations of the M dwarfs, we are able to obtain reddening values of the individual stars with very high accuracy. The typical uncertainty of our resultant reddening values is only 0.03\,mag in \ebr. We have also derived the reddening coefficient \rbr, as a function of the stellar intrinsic colour \bri\ and the reddening values \ebv. Values of \ebv, converted from \ebr, are also listed in our catalogue. Our derived reddening values are in good agreement with those deduced from the previous works.

Our catalogue contains high precision reddening values of a large sample of stars in the solar neighbourhood. The catalogue, combined with the robust distances from the Gaia parallaxes, will be important basic data for us to study the fine structure and properties of the interstellar dust nearby. 

\section*{Acknowledgements}
This work is partially supported by the National Key R\&D Program of China No. 2019YFA0405500, National Natural Science Foundation of China 12173034 and 11833006, and Yunnan University grant No.~C619300A034. We acknowledge the science research grants from the China Manned Space Project with NO.\,CMS-CSST-2021-A09, CMS-CSST-2021-A08 and CMS-CSST-2021-B03. 
This research made use of the cross-match service provided by CDS, Strasbourg.

Guoshoujing Telescope (the Large Sky Area Multi-Object Fiber Spectroscopic Telescope LAMOST) is a National Major Scientific Project built by the Chinese Academy of Sciences. Funding for the project has been provided by the National Development and Reform Commission. LAMOST is operated and managed by the National Astronomical Observatories, Chinese Academy of Sciences.

This work presents results from the European Space Agency (ESA) space mission Gaia. Gaia data are being processed by the Gaia Data Processing and Analysis Consortium (DPAC). Funding for the DPAC is provided by national institutions, in particular the institutions participating in the Gaia MultiLateral Agreement (MLA). The Gaia mission website is https://www.cosmos.esa.int/gaia. The Gaia archive website is https://archives.esac.esa.int/gaia.

\section*{Data availability}

The data underlying this article is available online via ``\url{http://paperdata.china-vo.org/diskec/lstmdwarfs/table1.fits}''.

\bibliographystyle{mnras}
\bibliography{ref}

\end{document}